\documentclass[fleqn,10pt]{wlscirep}
\usepackage[utf8]{inputenc}
\usepackage[T1]{fontenc}
\usepackage{float}

\title{Relationship between ideology and language in the Catalan independence context}

\author[1]{Julia Atienza-Barthelemy}
\author[1]{Samuel Martin-Gutierrez}
\author[1]{Juan C. Losada}
\author[1,*]{Rosa M. Benito}
\affil[1]{Grupo de Sistemas Complejos, Escuela Técnica Superior de Ingeniería Agronómica, Alimentaria y de Biosistemas, Universidad Politécnica de Madrid, Av. Puerta de Hierro, 2, 28040 Madrid, Spain.}

\affil[*]{To whom correspondence should be addressed. E-mail: rosamaria.benito@upm.es}

\begin{abstract}
Political polarization generates strong effects on society, driving controversial debates and influencing the institutions. Territorial disputes are one of the most important polarized scenarios and have been consistently related to the use of language. In this work, we analyzed the opinion and language distributions through Twitter data of a particular territorial dispute around the independence of Catalonia. We infer a continuous opinion distribution by applying a model based on retweet interactions, previously detecting elite users with fixed and antagonist opinions. The resulting distribution presents a mainly bimodal behavior with an intermediate third pole that shows a less polarized society with the presence of not only antagonist opinions. We find that the more active, engaged and influential users hold more extreme positions. Also we prove that there is a clear relationship between political positions and the use of language, showing that against independence users speak mainly Spanish while pro-independence users speak Catalan and Spanish almost indistinctly. However, the third pole, closer in political opinion to the pro-independence pole, behaves similarly to the against-independence one concerning the use of language. 

\end{abstract}
\begin{document}
	
	\flushbottom
	\maketitle
	\thispagestyle{empty}

\section*{Introduction}

In many political scenarios, public opinion is often divided into two extreme and opposite positions. In sociological terms, this process is called polarization, and has important social consequences 
%widely studied 
\cite{epstein2007polarized, persily2015solutions}. Some of them may affect the debate dramatically, because "A highly polarized system presumably produces clearer party choices, stimulates participation, affects representation, and has more intense partisan competition" \cite{dalton2008quantity}. Others strongly influence the actions of the government, like the gridlock effect, which entails a higher difficulty to pass laws \cite{binder2000going}. Furthermore, in recent times acute polarization has emerged around key social topics, like Brexit \cite{hobolt2018divided} and other controversial matters subjected to public referendum \cite{bbc_news_2016, refturk}. Given the significance of the subjects that generate this social phenomenon and its ubiquity, political polarization has been extensively studied. Some relevant examples are polarization of media \cite{bernhardt2008political, prior2013media}, Twitter \cite{conover2011political}, climate change people's position \cite{antonio2011unbearable}, American public \cite{fiorina2008political} and Venezuela \cite{morales2015measuring}.

In order to get insight into the main features of a polarized context, it is often useful to analyze the way people interact and debate. In many significant scenarios, the differences in the use of language or even the use of different languages is key to understand the mechanisms that underly polarization. After all, it is well known that the use of different languages result in different cognitive mechanisms in the people who use them \cite{suszczynska1999apologizing, goldin2008natural}. Moreover, languages are often powerful symbols of social categories as age \cite{montrul2008gender}, ethnicity \cite{michelle2010languages}, gender \cite{liladhar1998talking, shin2013social} and social class \cite{shin2013social}.

In this work, we are interested in the polarized system around the Catalan independence issue. Catalonia is an autonomous region in the northeastern of Spain where there is an important social and political movement bidding for independence. From the 17th century this territorial conflict has intermittently dominated Spanish affairs. In the 2010s, support for Catalan independence has grown, leading to an increase in the regional parliamentary representation of pro-independence parties: Partit Demòcrata Europeu Català (Catalan European Democratic Party) or PdeCat, Esquerra Republicana de Catalunya (Republican Left of Catalonia) or ERC and Candidatura d'Unitat Popular (Popular Unity Candidacy) or CUP. On the other hand, there is a significant presence of parties calling for the 'indisputable' Spanish unity: Partido Popular (People's Party) or PP, Ciudadanos (Citizens) or C's and Partit dels Socialistes de Catalunya (Socialists' Party of Catalonia) or PSC, associated to the Partido Socialista Obrero Español (Spanish Socialist Workers' Party) or PSOE. Although the majority of parties in Catalonia defend one of these two positions, there are parties with a slightly more intermediate opinion; for example, Podemos (We can).

The Catalan independence issue has lately dominated Spanish institutional political debates. During the time interval studied here, a coalition of pro-independence parties held a parliamentary majority, and "On November 9, 2015, the Catalan parliament narrowly approved a measure to implement a 'peaceful disconnection from the Spanish state.' Rajoy [the Spanish Prime Minister at the time] immediately reiterated the central government’s position that any such move would be illegal and opposed by Madrid."\cite{rodriguez_2019}. A highlight of this measure is the promise to hold a referendum which, despite the opposition by the central Government, was carried out on October 1, 2017. ``Catalan officials stated that turnout was around 42 percent, with 90 percent of voters voicing their support for independence'' \cite{rodriguez_2019}. This process of disconnection of the Spanish state is commonly known as 'procés'.

Our goal is measuring how this opinion is distributed among the general population in order to see if this polarization has been transferred from the  parties' elites to the remainder population, as some sociological studies affirm \cite{barrio2017reducing}. For this purpose, we will use data of the social networking site Twitter. In the Internet era, people have changed the way they express their ideas and emotions. Twitter is a platform where millions of users are actively posting their opinion and interacting with each other. Twitter is highly accessible and the amount of data it generates allows us to infer information that may be a reflection of reality. In this work we will infer the opinion of users that participate in a conversation about the Catalan independence issue based on their interactions. It is worth pointing out that approximately half of the Twitter users in Spain took part in this conversation \cite{statista}. We choose the Twitter conversation of the time interval from 15/09/2017 to 03/11/2017 because it covers several key off-line events related to the "procés": 

\begin{itemize}
	\item On the 1st of October 2017, a referendum was held on the question of independence, without approval of the central Government.
	\item On October 10th, the Catalan Government carried out the declaration and suspension of independence of Catalonia:  "'I assume the mandate of the people for Catalonia to become an independent state in the shape of a republic,' Mr. Puigdemont [the President of the Government of Catalonia at the time] said, before adding, seconds later, that he and his Government would 'ask Parliament to suspend the effects of the declaration of independence so that in the coming weeks we can undertake a dialogue.'" \cite{minder_kingsley_2017}. 
	\item On the 27th of October, the Catalan parliament declared, without the Spanish Government agreement, the independence of Catalonia (unilateral declaration of independence or DUI by its initials in Spanish). The response of Spanish Government was the enforcement of direct rule by the Spanish Government through the use of Article 155 of the Spanish Constitution \cite{ce}.
\end{itemize}

\begin{figure}
	\centering
	\includegraphics[width=0.5\textwidth]{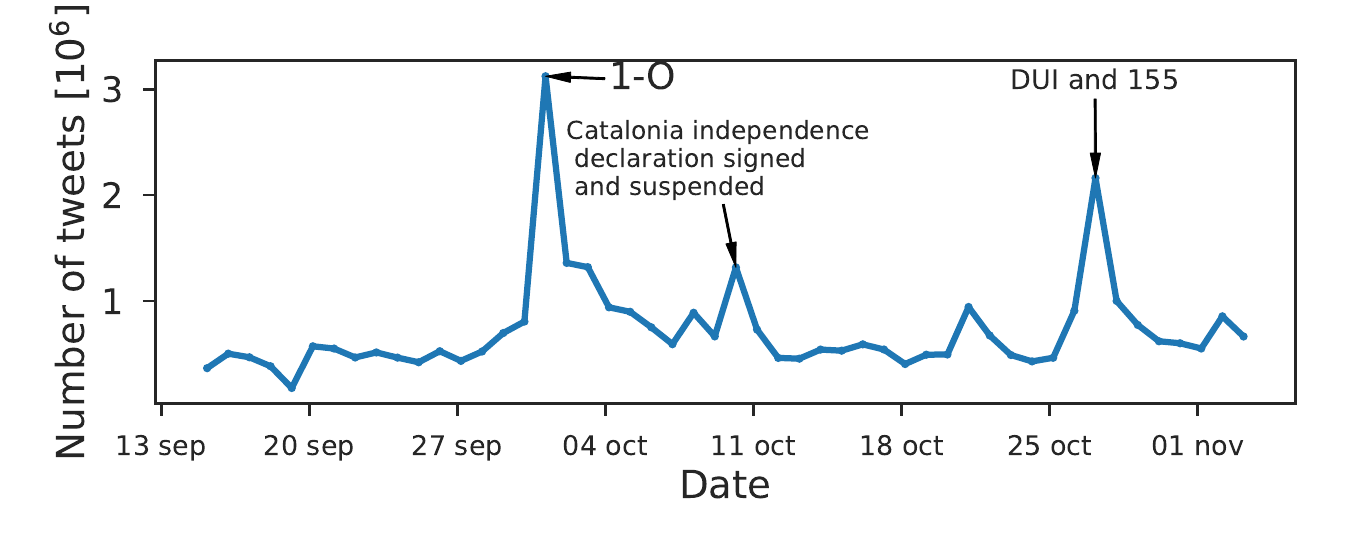}
	\caption{\small Temporal evolution of the number of tweets published in a Twitter conversation about the Catalan independence issue from 09/11/2017 to 11/04/2017. The three highest activity peaks match the off-line events of: the referendum day (1-0), the Catalonia independence declaration signed and suspended on the same day, and the Unilateral Declaration of Independence (DUI) and the enforcement of the Article 155 of the Spanish Constitution.}
	\label{fig:timeseries}
\end{figure}

The activity time series (number of tweets per day) is shown in Figure  \ref{fig:timeseries}. The referendum day, the 1st of October, is the highest peak of the time series. The declaration and suspension of independence matches the third highest peak on the 10th of October. The second highest peak corresponds with the DUI and the enforcement of Article 155 on the 27th of October. The average number of tweets per user each day remains considerably constant on three tweets per user. This means that participation peaks are due to new users joining the conversation rather than users writing more tweets, a result consistent with previous Twitter studies \cite{martin2018recurrent}.

In Catalonia there are two co-official languages: Catalan and Spanish. In line with the discussion above, sociological studies have consistently linked territorial disputes and the use of different languages \cite{inglehart1967language, shabad1982language, craith2011politics}. Following this idea, we analyze how the use of language is related to political opinion in the polarized system under study, that is, we analyze the use of Catalan and Spanish in relation to the opinion about the Catalan independence. To the best of our knowledge, every study so far has measured the language-ideology relationship using discrete values. In this work, we adopt a methodology that enables us to infer political opinion, measure the use of language and analyze their interplay in a approximately continuous way. Furthermore, while previous works present aggregated results, we carry out this study with finer granularity by performing the analysis at the user level. This approach allows to unravel more complex patterns.

The methodology adopted to infer the opinion of the Twitter users is based on the work by Morales et al. around the polarization in Venezuela \cite{morales2015measuring}. To this end, we first build the retweet network of the conversation and developed a methodology to choose the set of elite users with fixed and antagonist opinions that will propagate their opinion through the network. The opinion of the rest of users is inferred by iteratively averaging the opinions of their neighbors by means of a DeGroot process. Then, the opinion distribution is characterized with the polarization index and its underlying parameters \cite{morales2015measuring}. On the other hand, a language index has been defined according to the number of tweets written in each language by a given user. These computations yield a pair of values for each user that are used to study the interplay between opinion and language.

\section*{Results and discussion}
\paragraph{}
Here we present the results got from the analysis of the opinion distribution, the language distribution and the interplay between ideology and use of language. At the same time, we analyze and discuss the main conclusions that we can extract from these results.

\subsection*{Opinion analysis}
		
In this section we study the opinion distribution resulting from applying the model to estimate users opinions. Besides, we analyze the characteristics of this distribution for active, influential and engaged users, and how the distribution evolves in a daily base. 
	
As we explain in the Data and Methods section, in order to infer the users' opinion, we first analyze their interactions. Among them, we choose the retweet interaction because it is a broadcasting mechanism \cite{yaqub2017analysis, martin2018recurrent, borondo2012characterizing} that usually implies that the user back up the original tweet. We build retweet networks and adopt a model \cite{morales2015measuring} based on the De Groot process to infer the opinion of a user in a network as the average of her neighbor's opinions. This model requires to define two sets of users: users with a fixed opinion, called elite, and the users whose opinion is to be inferred and computed, called listeners. We choose highly influentials and engaged users, i.e., users with a high number of retweets that participate a high number of days, and we study the community structure of these users. The elite users are made up by the two communities that include the largest number of users where all politicians contained are from political parties completely in favor of the Catalan independence, or completely against it, respectively. The rest of users, called Listeners users, must be "connected" to the elite in the retweet network, which means that there must be a directed path that starts in each listener and ends in at least one elite user.

\begin{figure}[H]
	\centering
	\includegraphics[width=0.7\textwidth]{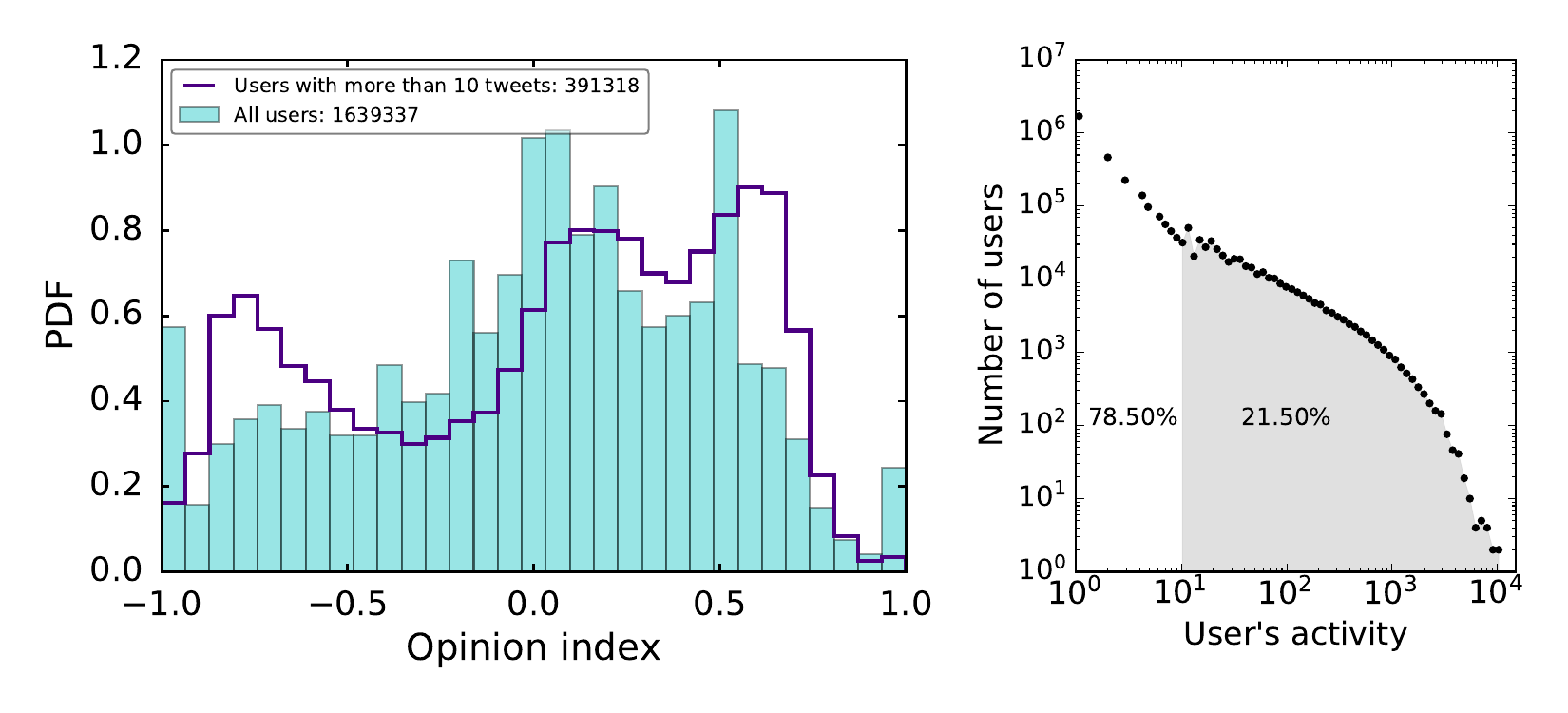}
	\caption{ \small Left: opinion distribution of all the users and of the users with an activity corresponding to more than 10 tweets in the Catalan independence conversation.  Right: distribution of user activity. Grey area shows the percentage of users that have posted less than 10 tweets in the whole period.}
	\label{fig: opinion_distribution}
\end{figure}

The considered network is the aggregated retweet network corresponding to the listeners and elite users that tweeted from 09/15/2017 to 03/11/2017. The resulting opinion distribution when the opinion inference model is applied to this network is shown in the left of Figure \ref{fig: opinion_distribution}, where -1 means being completely against the Catalan independence and 1 being completely in favor. We have characterized it with the normalized distance between the poles $d$, their size difference $\Delta A$ and the polarization index $\mu$, which are defined in the Data and Methods section. The corresponding values are given in Table \ref{tab:opn_parameters}. We can see that the opinion is evenly distributed among the two poles $(55\%-45\%)$. On the other hand, as the gravity centers of each pole ((-1,0) and (0,1)) do not present very extreme values, the pole distance $d$ is only $0.39$. Accordingly, although $\Delta A$ is low $(0.10)$, the polarization index $\mu$ is fairly small $(0.35)$.

It is well known that in any social network there is a high number of users that hardly ever participate. We can see in the activity distribution, shown in the right of Figure \ref{fig: opinion_distribution}, that most users (78.50\%) have an activity smaller than 10 tweets (original tweets, retweets or quotes) on the 50 days time interval considered. As we estimate the users opinion based on their retweets connections, the resulting opinion distribution of a network with this activity distribution could be noisy. Therefore, if we only show the users with an activity greater or equal than 10 tweets, the resulting opinion distribution (left of Figure \ref{fig: opinion_distribution}) is a different version with 391318 users. It is noteworthy that, besides the pro-independences and against independence poles, we can infer the existence of a third pole, which has emerged spontaneously from the opinion inference process. The third mode of the distribution can be easily understood considering the existence of the political parties with alternative positions to being completely in favor or completely against the Catalan independence. This can also be explained from the results of the community analysis discussed in the "Model to estimate opinions" subsection of the "Data and Methods" section.

\begin{table}[H]
	\centering
	\scalebox{0.9}{
	\begin{tabular}{ | c | c | c | c | c | c | c | c |}
		\hline
		& $A^+$& $A_-$ & $\Delta A$ & $gc^+$& $gc_-$ & $d$  & $\mu$ \\ \hline 
		All users & 0.55 & 0.45 & 0.10 & 0.36 & -0.42 & 0.39 & 0.35\\ \hline 
		Users with more than 10 tweets & 0.59 & 0.41 & 0.18 & 0.39 & -0.50 & 0.45 & 0.37 \\ \hline 
	\end{tabular}}
	\caption{\label{tab:opn_parameters} \small Parameters, defined in Data and Methods, that characterize the opinion distributions of Figure \ref{fig: opinion_distribution}. }
\end{table}

\subsubsection*{Active, influential and engaged users}
\paragraph{}
	
In the previous section, we filtered out users with low activity in order to obtain a less noisy opinion distribution. A complementary approach is to analyze those users that participate at least a given number of days in the conversation. Users that participate a high number of days are called engaged users. Another relevant subset of users worth of analyzing are those with a high number of retweets received, called the influential users.

\begin{figure}
	\centering
	\includegraphics[width=0.45\textwidth]{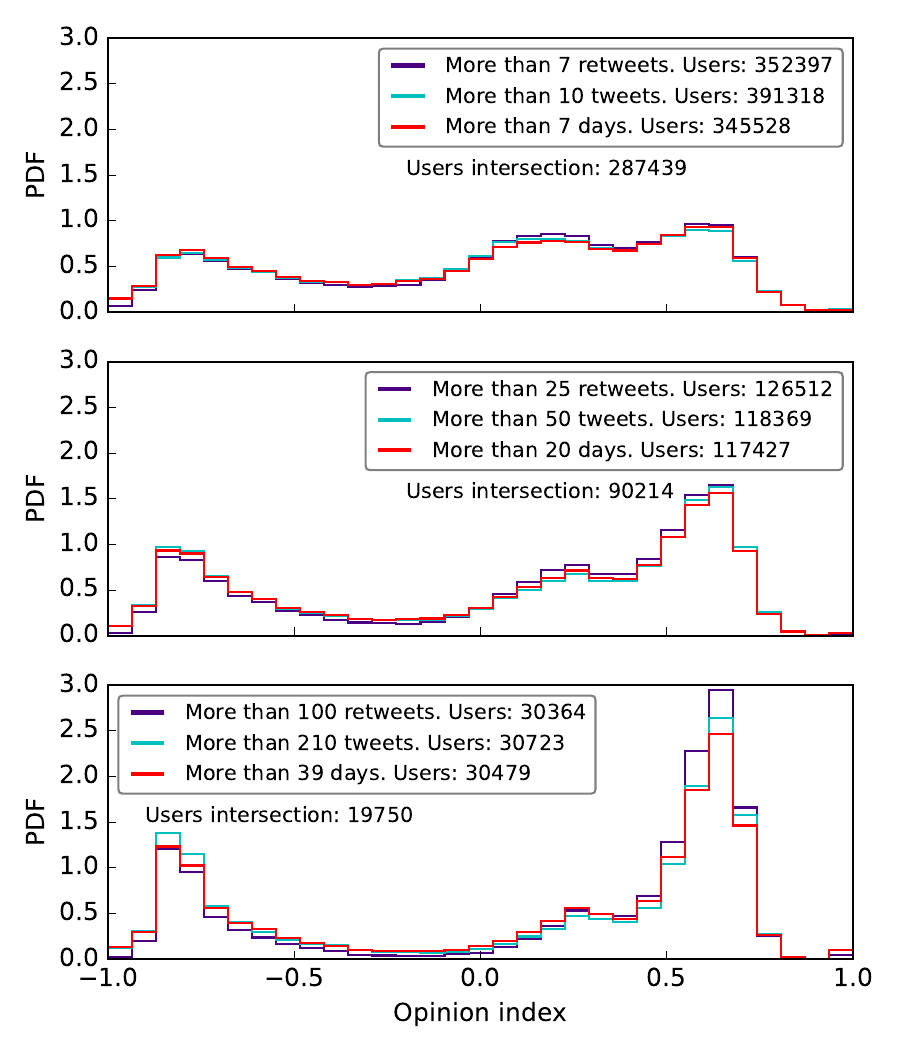}
	\caption{ \small Comparison of the effect of activity, influence and engagement on the users' opinion distribution in a Catalan independences Twitter conversation. Results are filtered by activity (blue), retweets received (purple) and days of participation (pink) for three different threshold values increasing from top to bottom panels.}
	\label{fig:type_users}
\end{figure} 
		
In order to measure the effect of the most active, engaged and influential users, we apply these three filters so that each of the three subsets contains approximately the same number of users. If we consider the activity filter of 10 tweets, the influence filter of having received at least 7 retweets and the engagement filter of 7 days, the size of the subsets are 391318, 352397 and 345528 users, respectively. Besides these thresholds, we consider two more increasingly restrictive (i.e. higher) values for activity, retweets and participation. The resulting distributions for each threshold are shown in Figure \ref{fig:type_users}. The first thing to notice is that the shapes of the distributions of the same restriction level are very similar. In fact, the intersection between the active, influential and engaged users on each threshold is very high, particularly in the less restrictive level, meaning that the sets of users are almost the same.

Furthermore, it can be seen that as we increase the threshold (from top to bottom in Figure \ref{fig:type_users}), the extreme poles get progressively narrower while the central pole approximately maintains its shape, with a slight decrease in size and height. Besides, the pro-independence pole increases so that in the last threshold the positive part of the distribution contains 67\% of the users. That is, the more active (engaged or influential) a user is, the more probable it is that she is found in an extreme position in the ideological spectrum. Moreover, the progressive narrowing of the distribution around the three modes (the three poles) means that opinion variability significantly decreases for the most active users. This effect has been corroborated by computing the entropy of the distributions of active users for each of the three thresholds considered and verifying that, as the threshold increases, the entropy of the distribution decreases. The values of the entropy for each activity threshold, calculated as described in the Supplementary Information, are $h_{10}=0.54, h_{50}=0.37$ and  $h_{210}=0.08$.
	
\subsubsection*{Daily analysis}
\paragraph{}
	
\begin{figure}
	\centering
	\includegraphics[width=0.6 \textwidth]{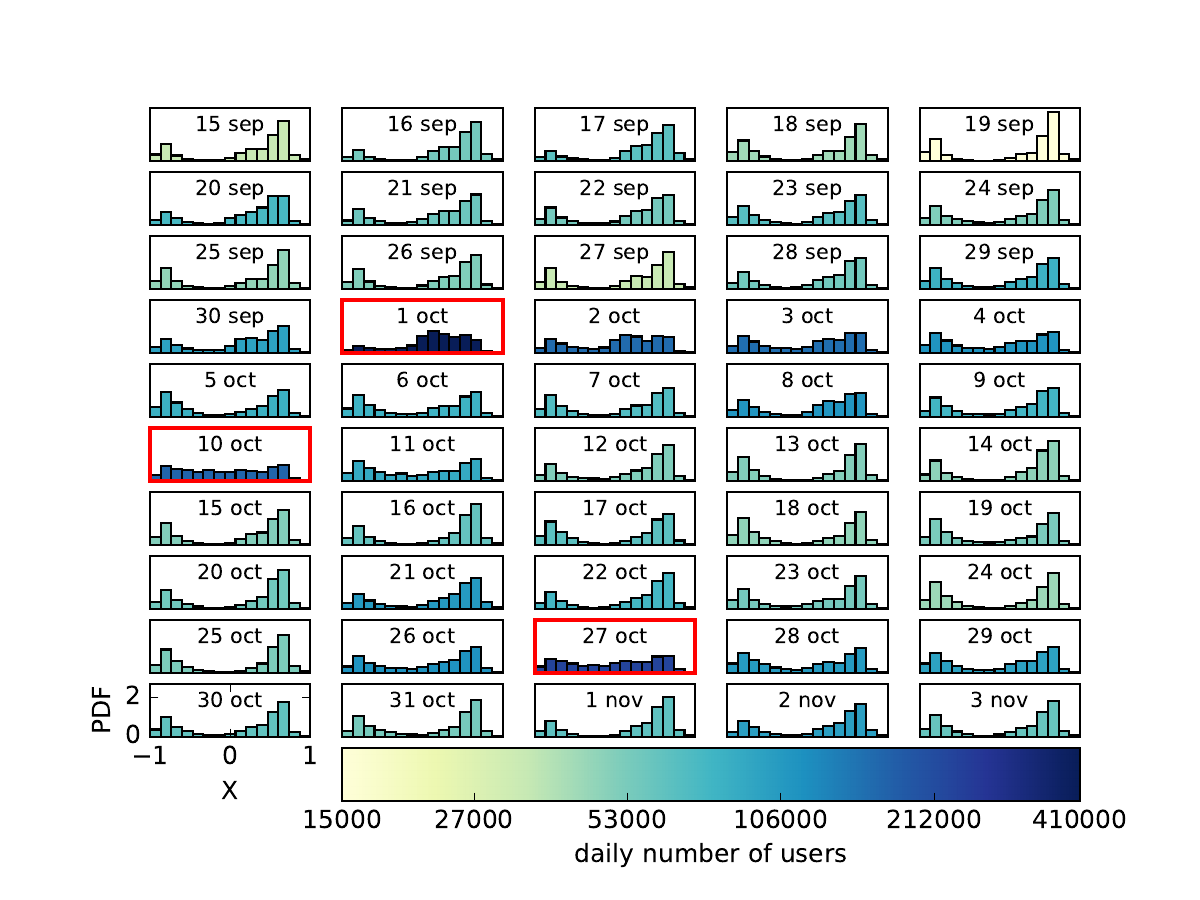}
	\caption{ \small Time evolution of the daily distributions of opinion index ($X_i$) for the Catalan independence twitter conversation. Color indicates the number of users that participate each day. The three days framed in red correspond to the three activity peaks observed in Figure \ref{fig:timeseries}.}
	\label{fig:dist_days}
\end{figure}

As we discussed in the introduction, there are some off-line relevant events that correspond to bursts of activity (see Figure \ref{fig:timeseries}) generated by the incorporation of users to the conversation. We want to study how this behavior relates to the opinion distribution of the users that participate each day. For this purpose, we got the users' opinion computed with the aggregated retweet network of the entire time interval, i.e., the network resulting from considering all retweets that took place during the 50 days. With these computed opinions, we build the opinion distribution of each day selecting only the users that participated that day. The resulting daily opinion distributions are shown in Figure \ref{fig:dist_days}.  
	
It can be noticed that during the first days, the opinions are mostly distributed between the two extremes values, the right one with a larger number of users. When the day of the referendum arrives, the opinion distribution becomes more diverse and this greater diversity remains until October 5, when the opinion returns to be distributed mainly on the two extremes values. On the 8th of October the diversity increases again until the 11th. Afterwards, it returns to be restricted to the two extremes until the days around the 27th of October. In the last days the opinion is distributed again at both extremes. Overall, it seems that the greater the participation the greater the diversity of opinions. Also, the positive side of the opinion index seems to have more users than the negative one. 
	
\begin{figure}
	\centering
	\includegraphics[width=0.55\textwidth]{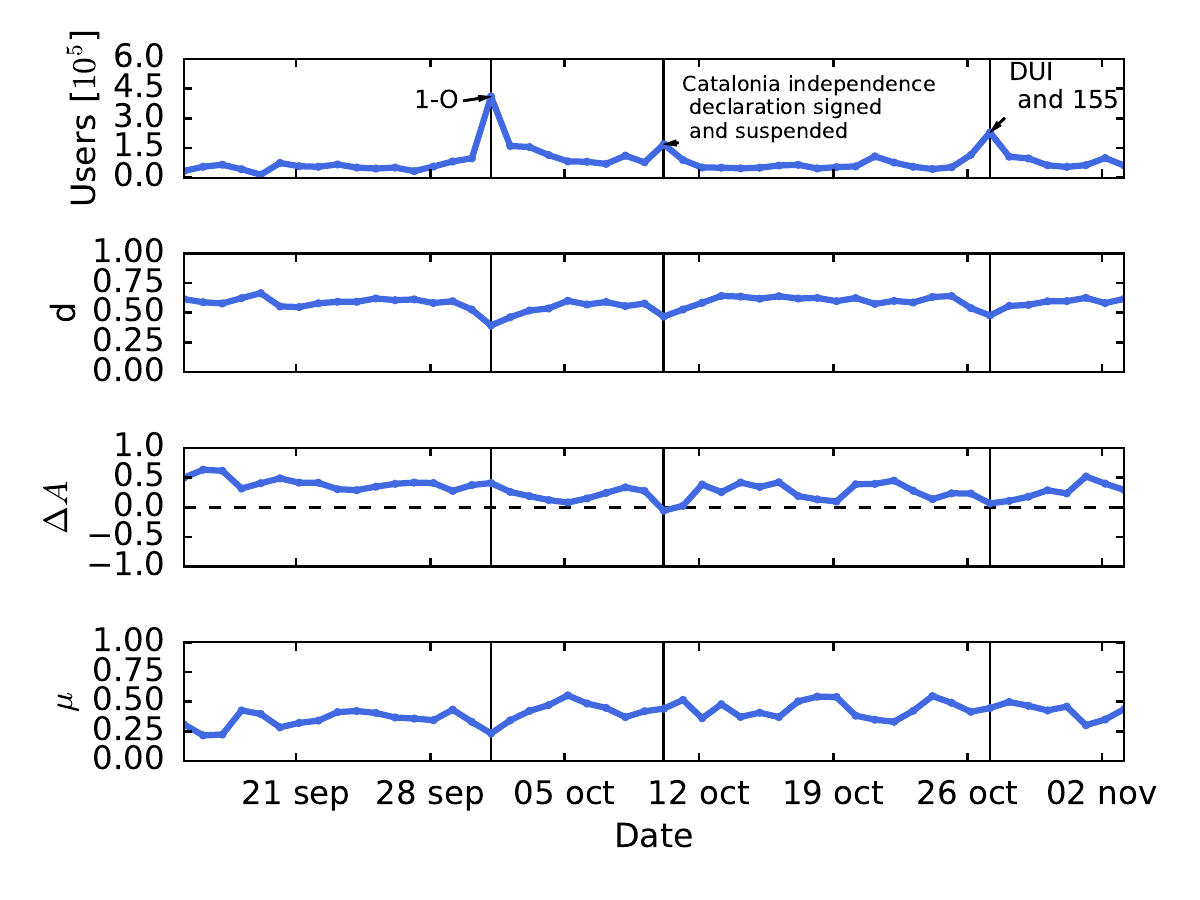}
	\caption{ \small Temporal evolution of the number of users and the normalized pole distance (d), relative population size ($\Delta A$) and the polarization index $\mu$ for the Catalan independence conversation.}
	\label{fig:par_days}
\end{figure} 

In order to measure and quantify these observations about the shape of the distribution, we calculate three parameters of each daily distribution: the pole distance $(d)$, the normalized difference in population sizes ($\Delta A$) and the polarization index ($\mu$). The temporal evolution of these parameters are shown in Figure \ref{fig:par_days}. In this figure we see that $d$ remains relatively high ($\sim 0.6$) and remains at an almost constant value excluding the days with very high activity, when the distributions are broader and the value $d$ decreases. On the other hand, $\Delta A$ is always positive, i.e., the pro-independence pole is always greater than the against independence pole. However, $\Delta A$ is more positive at the beginning and then the sizes of the two poles become more similar. Finally, $\mu$, as it is a combination of these two parameters and $d$ is roughly constant, is dominated by the fluctuations of $\Delta A$. Although pole distance $d$ is quite high, due to the fact that most of users are located in one of the two poles, the measure of polarization $\mu$ did not correspond to a high polarization.
	
\subsection*{Interplay between opinion and language}
\paragraph{}
	
So far, we have studied the political opinion of the users of the Catalan independence conversation. However, as we have mentioned, the use of language is often related to social and political categories, and it has been proven that underlies many territorial conflicts. Taking into account that there are two co-official languages in Catalonia, it is of interest to study the relationship between the opinion and the language used by the users in their tweets.

\subsubsection*{Language distribution}
\begin{figure}[h]
	\centering
	\includegraphics[width=0.55\textwidth]{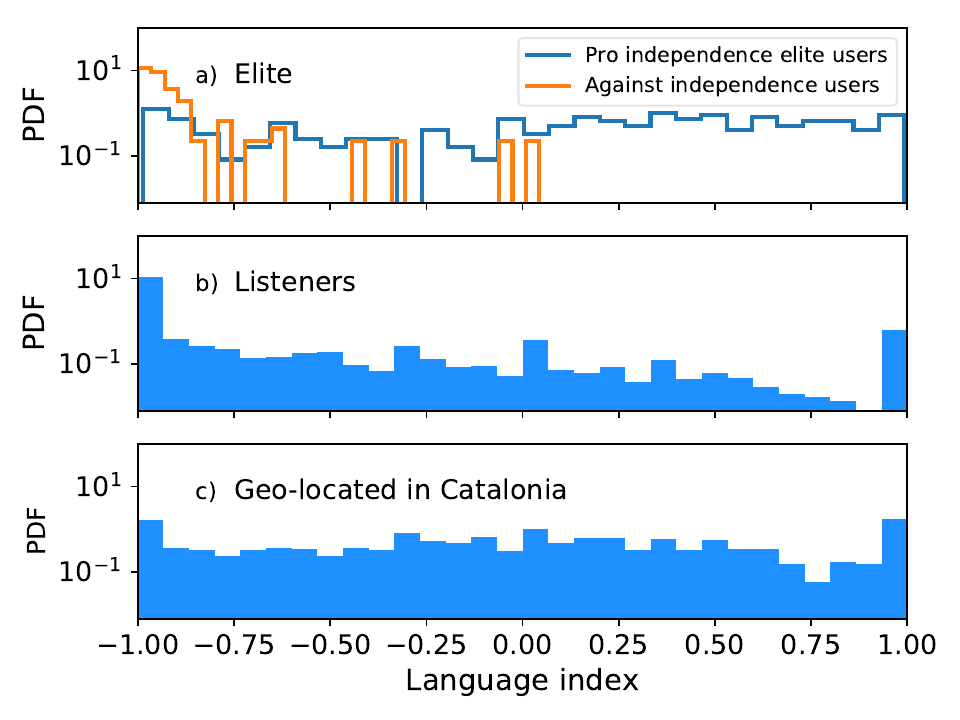}
	\caption{ \small Distributions of language index (L) for (a) Elite users, where the pro-independence elite users are in blue and the against-independence in orange, (b) Listeners users and (c) Geo-located in Catalonia users.}
	\label{fig:lang_dis}
\end{figure}

\paragraph{}
First of all, we compute the Language index, $L_i$, of the users in the retweet network, as described in the Data and Methods section. This calculation has been carried out taking into account all the users' tweets (original, retweets and quotes) published during the 50 days time interval. The $L$ index distributions for the elite users (top), listener users (middle) and users whose tweets were geolocated in Catalonia (bottom) are shown in Figure \ref{fig:lang_dis}.

The language distribution of the Listeners shows a clear large majority of users who write (or retweet or quote) only in Spanish (notice that the vertical axis is in logarithmic scale). This makes intuitive sense because, although the conversation is about Catalonia, the subject has been of general interest in Spain. However, the language distribution of the against independence elite users has a high peak of users that speak only Spanish, while the pro independence elite users have an approximately uniform distribution. Finally, the language distribution of Catalan users have an approximately uniform distribution too.

\subsubsection*{Language and opinion interplay}
\paragraph{}

\begin{figure}
	\centering
	\includegraphics[width=0.55\textwidth]{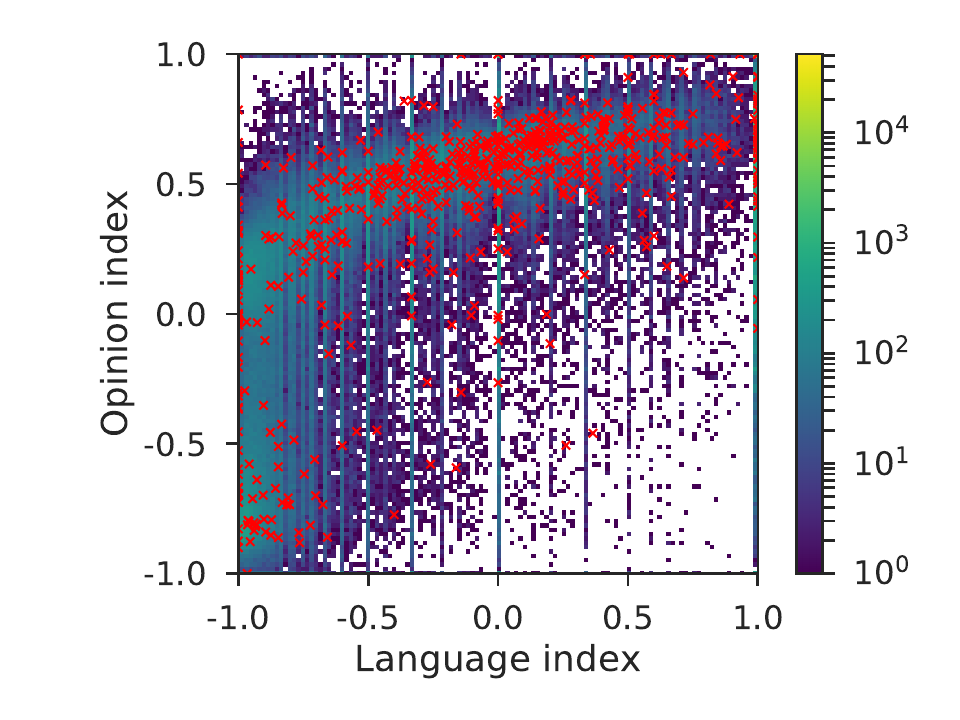}
	\caption{ \small Interplay between language index and opinion index of the users of the Catalan independence Twitter conversation. Color measures the number of users located in the 2D bin. The red x's correspond to Geo-located in Catalonia users.}
	\label{fig:lang_opinion_dis}
\end{figure}

Since each user has been assigned with an opinion value and a language index, we have represented the relationship between ideology and language use by plotting these couples of values in Figure \ref{fig:lang_opinion_dis}, where color indicates density of users and the red x's correspond to individual users geolocated in Catalonia. In order to study this relationship, we have measured the Spearman correlation coefficient between the language and opinion indices and the value obtained is 0.46. If users used Catalan or Spanish regardless of their political position, the distribution would be homogeneous, yielding a value of $\sim0$ for the Spearman correlation. Notice that if only users of the independence pole (opinion index from 0 to 1) used mainly Catalan, and users of the against independence pole (opinion from $-1$ to $0$) used mainly Spanish, the upper-left and lower-right quadrant of the 2D histogram would be approximately empty, which is not the case either. 

In more detail, what we see in Figure \ref{fig:lang_opinion_dis} is that users with an opinion index from $-1$ to $0.3$ (from completely against-independence to slightly pro-independence) have a language index close to $-1$, i.e., they speak almost exclusively Spanish. On the other hand, for users with an opinion index between $0.3$ and $1$, the possible language index values are wider and there exists a monotonously increasing tendency to use more preferably Catalan as the opinion gets closer to the pro-independence pole. The Catalan geo-located users, the red x's, follow an analogous behavior.

\subsubsection*{Active, influential and engaged users}
\paragraph{}

\begin{figure}
	\centering
	\includegraphics[width=0.8\textwidth]{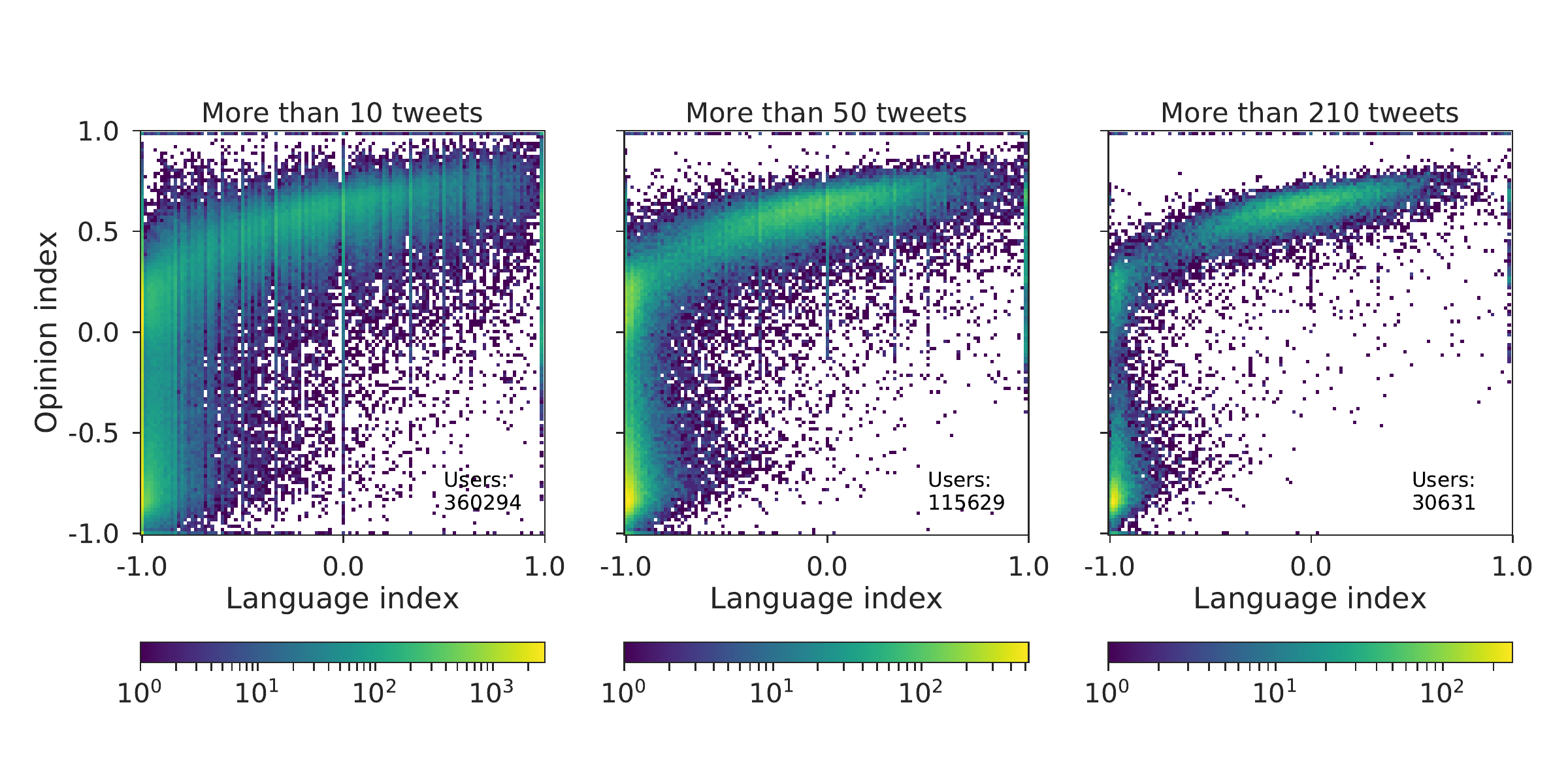}
	\caption{Effect of the activity in the interplay between language index and opinion index. Three different activity thresholds have been used to filter users' activity: 10 tweets (left), 50 tweets (center) and 210 tweets (right). Color indicates the number of users located in each 2D bin. }
	\label{fig:lang_opinion_filter}
\end{figure}

As we discussed above, the opinion inference for users with lower activity is poorer and more noisy. The same applies for the language index. Hence, we have performed an analogous filtering of the users and kept only those with higher activity in order to obtain clearer results. Since we also verified that the intersection between active, influential and engaged users is quite high, the results for the active users will be analogous as those obtained for the engaged and influential users. Therefore, to avoid redundancies, we present here only the results of the effect of the active users. In Figure \ref{fig:lang_opinion_filter} we show the ideology-language interplay when the users are filtered using the same increasingly restrictive thresholds of activity that were employed before in Figure \ref{fig:type_users}. Here we see that the distribution follows the same pattern for the three values of the threshold considered, but it becomes narrower as its value is increased. In addition, there are three areas that stand out because of the higher density of users on them. These areas correspond to the three poles observed in Figure \ref{fig:type_users}. Thus, the users of the pole with an opinion index around $-0.7$ (against independence) and the pole with the most central opinion index speak almost exclusively Spanish. Conversely, the users of the pole with an opinion index around $0.6$ (pro-independence) speak in a wide range between Catalan and Spanish. That is, although the {\it third} pole (the central one) is closer to the pro-independence pole in the opinion index, it behaves similarly to the against independence pole in the use of language.

\section*{Conclusions}

Our main goal in this work was to infer the opinion of the population in a polarized system and study its relation with the use of the two co-official languages spoken in Catalonia. The polarized system studied is centered around the Catalan independence issue and has been analyzed through a Twitter dataset. The dataset is made up of tweets about the topic published in the period between 09/15/2017 and 03/11/2017. During this period important events occurred, the most relevant being the celebration of a referendum on independence not approved by the Spanish Government. Some of these events were clearly reflected as activity peaks in the time series of the conversation.

The users’ opinion has been inferred by applying a model to estimate opinions \cite{morales2015measuring} based on their retweet interactions. When the model is applied to the retweet network aggregated for the whole 50 days period, we obtain an opinion distributed evenly between the two poles (45\% -55\%) and a relatively low pole distance ($d=0.39$) that leads to a low polarization index $(\mu=0.35)$. We have applied a user filter  to focus on {\it active} users (those with more than 10 published tweets) to improve the reliability of the opinion inference, obtaining a different opinion distribution. In particular, the independence pole now covers 62\% of the users and a {\it third pole} is detected in a moderate/pro-independence position. Notice that the elite users (the opinion seeds) are assigned fixed and extreme opinions, which implies that the proposed methodology is able to reveal emergent intermediate positions in apparently polarized contexts. In this case, the third pole can be explained by the existence of political parties that defend a middle ground.

To further explore the effect of the activity filter, we have applied two increasingly higher activity thresholds and analyzed how the opinion distribution changes for the resulting subset of users. Complementarily, we have also extracted subsets of users that participate a high proportion of days ({\it engaged} users) and users that received a large number of retweets ({\it influential} users) in order to gain more insight about how the opinion distribution is shaped depending on the user behavior.
By doing this, we have found that active, influential and engaged users are practically the same subset of users. Besides, as the thresholds of the three filters are increased, the extreme poles become narrower, the pro-independence pole increases in size and height and the central pole approximately maintains its shape with a slight decrease in size and height. That is, the more active (engaged or influential) a user is, the more probable it is that she is found in an extreme position in the ideological spectrum. Moreover, the progressive narrowing of the distributions around the three modes (the three poles) means that opinion variability significantly decreases for the most active users. This has been corroborated by computing the entropy of the distributions of active users for each of the three thresholds and verifying that, as the threshold increases, the entropy of the distribution decreases, with a difference of one order of magnitude between extreme values of the threshold.

Since the bursts of activity of the Twitter conversation can be attributed to the incorporation of new users corresponding to relevant off-line events, we have studied how the daily distribution of opinion changes when we consider only the users that participated a given day. We have seen that, whenever the number of users is low, the daily distributions are mostly bimodal with most of the users concentrated on the two extremes, the right one with a larger number of users. However, when the number of users increases, the opinion distribution becomes more diverse. In order to measure these observations, the pole distance $(d)$, the normalized difference in population size ($\Delta A$) and the polarization index ($\mu$) of each day have been calculated. We have seen that $d$ remains relatively high during all the time interval considered, i.e., the poles are far apart, but it decreases around the bursts of activity. Furthermore, $\Delta A$ stays always positive, that is, the pro independence pole is always greater than the against independence pole. However, the first days $\Delta A$ is higher and then it begins to decrease, so the sizes of the two poles become more similar. Finally, since $d$ remains almost constant, $\mu$ is dominated by the $\Delta A$ fluctuations.

There are two co-official languages in Catalonia: Catalan and Spanish. It has been proven that, as in many other social categories, the use of language in the context of a territorial conflict is related to certain political positions. For this reason, we have studied the interplay between the inferred political opinion and the language used. This use of language is quantified by the language index $L_i$, that measures the Spanish-Catalan usage ratio of each user. We have shown that there is a clear relationship between the political and the language indices, with a Spearman correlation of $0.46$. Users with an opinion index ranging from completely against-independence to slightly pro-independence have a language index close to $-1$, i.e., they speak almost exclusively Spanish. On the other hand, for users with an opinion index closer to the pro-independence pole, the possible language index values are wider. Furthermore, there exists a monotonously increasing tendency to preferably use Catalan as the opinion gets closer to the pro-independence pole. The Catalan geo-located users follow an analogous behavior. Besides, the pro-independence elite presents an approximately uniform language distribution while against-independence elite users speak mainly Spanish.

Finally, we have applied three different activity thresholds to the analysis of the opinion-language relationship. Although the global pattern remains similar for the three thresholds, the distribution becomes narrower when the threshold is increased. Moreover, there are three areas that stand out, corresponding to the three political poles. Users of the most against-independence pole (opinion index $\sim -0.7$) and users of the pole with the most central opinion index $(\sim 0.2)$ speak almost exclusively Spanish. Conversely, users of the pole with an opinion index nearer to the pro-independence extreme (opinion $\sim 0.3-1.0$) present a wider range of language use between Catalan and Spanish. In conclusion, although the {\it  third} pole (the central one) is closer to the pro-independence pole in political opinion, it behaves similarly to the against independence pole in the use of language.

Summarizing, our results show that the proposed methodology is able to reveal the complex patterns of the ideological landscape in a polarized context. It is worth emphasizing the detection of a third pole that naturally emerged from the application of the opinion inference model as well as the richness of the information extracted from the study of the interplay between language and ideology. Finally, we would like to point out that it is possible to carry out analogous analysis to those developed for the ideology-language relationship with other relevant social dimensions like income, gender, age, etc.

\section*{Data and Methods}
\paragraph{}
In this section we explain the data and methods used for the development of this study. In particular, we present the dataset, we discuss how the retweet network is built, the model to estimate opinions on this retweet network explaining the way elite's users are chosen, the parameters computed to characterize the opinion distribution and the method to detect and measure the language used by Twitter users.

\subsection{Data}
\paragraph{}
The data set employed for this study consists in Twitter messages, called tweets, about the Catalan independence issue. We have used the Twitter Streaming API to download tweets that contain at least one word of a set of keywords. It is known that this API provides all real time tweets matching the words filter if the data volume per unit time is small enough \cite{campan2018data}, as has been checked that is the case. The set of keywords used is:

1-O, 1O, procés, proces, procès, cataluña, catalunya, referendum, parlament, generalitat, catalans, catalanes, catalan, catalán, puigdemont, independencia, independència, urna, 155, dui.

We have included a very broad range of keywords as neutral as possible, both in Catalan and Spanish. The choice of keywords used to retrieve tweets completely determines what messages are obtained; therefore, depending on which keywords are chosen, a "content bias" could be induced on the results. However, due to the variety and neutrality of the terms employed, the possible bias will be minimized. At the time of the download, it was a burning issue between Spanish Twitter users. The resulting dataset is made up of 36090661 messages written by 2511319 users from 15/09/2017 to 04/11/2017.

\subsection{Retweet network}

\paragraph{}
We want to estimate the opinion of the users who participate in the conversation based on their interactions. Among the interactions that exist on Twitter (follow, retweet, quote, answer, etc), the retweet implies that the user agrees with the original tweet and she has enough interest to perform the retweet action. This is why retweets are used as proxy for influence \cite{yaqub2017analysis, martin2018recurrent, borondo2012characterizing}: whenever a user j retweets a message originally posted by the user i, j is being influenced by i's ideas and there is a link from j to i (Supplementary Fig.S3). This network is directed and weighted. Therefore, we will consider the retweet network built with the Twitter users who participated on the conversation as the social network in which we will infer the political opinion. We will use the aggregated retweet network of the time interval of the 50 days, i.e., the retweet network with all the retweets from 15/09/2017 to 03/11/2017.

\subsection*{Model to estimate opinions}

\paragraph{}
We use a model based on the De Groot process that estimates opinions of individuals in a continuous way, determined by their interactions in a network \cite{morales2015measuring}. The network used is the retweet network. In this model, two kinds of nodes are considered:
\begin{itemize}
	\item Elite: a minority of users with a fixed opinion that act like seeds of influence. We consider two sets of elite users with antagonist opinions, i.e., $X_s = -1$ or $X_s= 1$. 	
	\item Listeners: a majority of users that starts with an initial neutral opinion $X_l(0) = 0$. They will iteratively update their opinion value as the mean opinion of her incoming neighbors:
	\begin{equation}\label{eq:listeners}
	X_i(t) = \frac{\Sigma_jA_{ij}X_j(t-1)}{k_i^{out}}
	\end{equation}
	
	Where  $A_{ij}$ represents the elements of the networks adjacency matrix, which is 1 if and only if there is a link from j to i, and  $k_{i}^{out}$ corresponds to the outdegree of node i (the number of retweets made by i user). The outcome will be the opinion of all listeners users, with which we can build the opinion distribution of the system, lying in the range  $-1 \le X_l \le 1$.
\end{itemize}

\subsubsection*{Elite selection}
\paragraph{}
One of the main challenges of applying the model is the selection of the set of users that are part of the elite. Elite users are crucial since they are the seeds that will influence the opinion of the rest of users. Therefore, these users should be very influential and should have constant opinions. In addition, they should have antagonist opinions, so we need two type of elite users that represent completely in favor or completely against the Catalan independence positions. In this way, the full range of opinions can be inferred. We carry out the selection of the elite in two steps. 
	
\paragraph{First Step: }
Elite users must have a strong constant opinion because we will keep it fixed. A user that frequently participates in the conversation can be considered to be engaged in the subject and, consequently, to have a well defined opinion. The engagement of a user can be measured as the proportion of days (out of the total) that the user participates in the conversation. This is called the participation ratio. On the other hand, as the elite users will be seeds of influence, we need them to be relevant in the network.  Accordingly, they should be the most central users; for example, users with a high out degree. In other words, users with a large number of retweets received.
\begin{figure}
	\centering
	\includegraphics[width=0.5\textwidth]{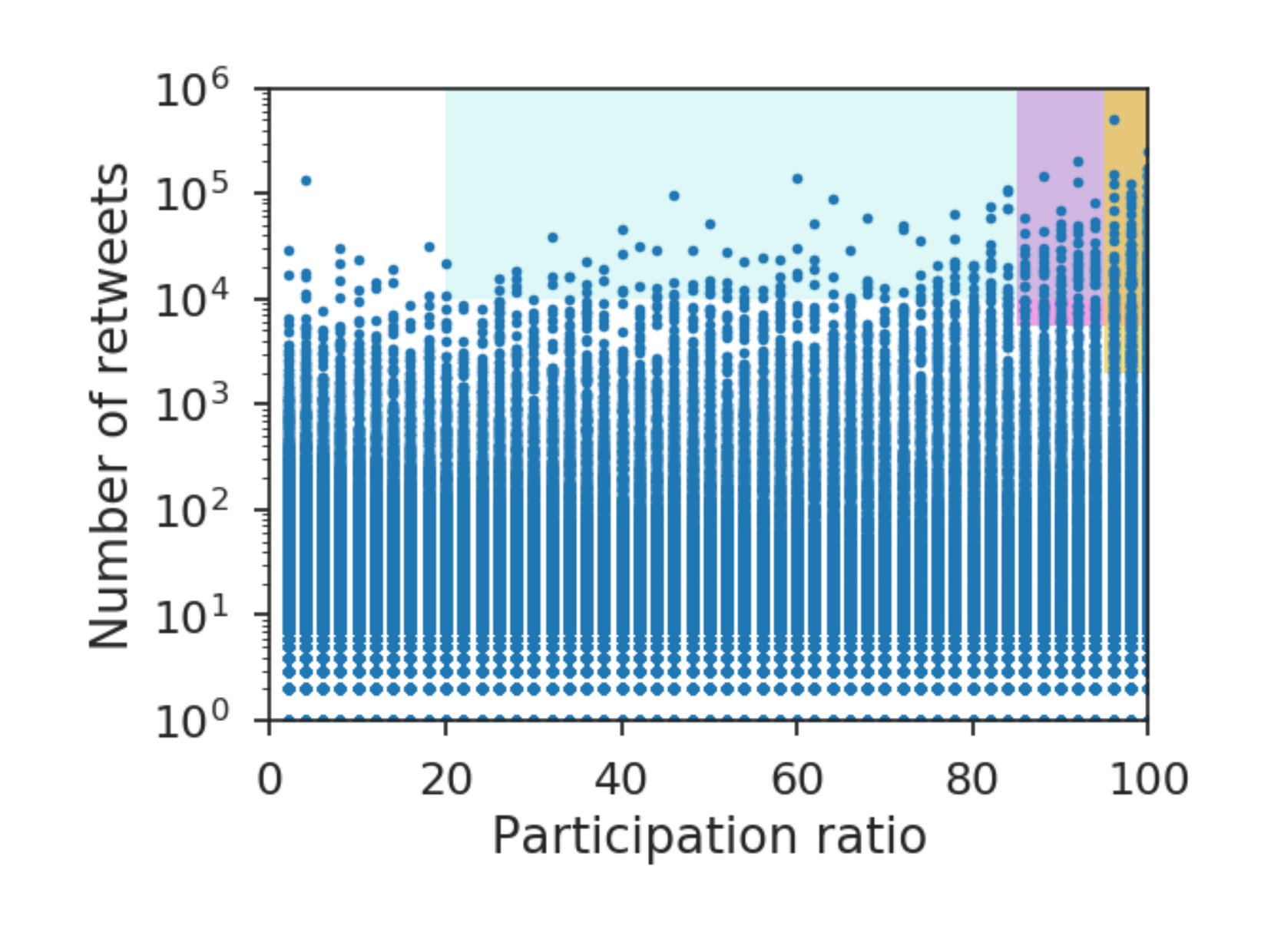}
	\caption{\small Retweets received vs. percentage of participation days of users in the Catalan independence Twitter conversation. Colored rectangles represent three subsets of users according to different criteria thresholds. We will work with the pink set, although the outcome is practically the same with the other two sets.}
	\label{fig:rectangles}
\end{figure} 
	
In Figure \ref{fig:rectangles}, the number of retweets versus the participation ratio of all users of the Twitter conversation about the Catalan independence issue is shown. Considering these two criteria, we pick out a set of users with a high participation ratio and a high number of retweets; specifically, users with a participation ratio > 80\% and a number of retweets > 5500. This users correspond, in Figure \ref{fig:rectangles}, to the subset marked as a pink rectangle. In order to verify the robustness of our results, we have also repeated these computations with two other user subsets. The yellow subset of Figure \ref{fig:rectangles} gives more importance to the participation ratio and the blue subset gives more importance to the number of retweets received. We have verified that the results are practically the same independently of the chosen users subset.
	
\paragraph{Second Step:}
In order to finally build the elite, once we have selected these highly engaged and influential users, we check which of them hold antagonist opinions to assign them the corresponding values $X_s=1$ or $X_s=-1$. In this context, antagonist opinions mean being completely in favor of the Catalan independence or completely against it. In order to apply this last criterion, we perform a community analysis for the whole retweet network by the Nested Stochastic Block Model \cite{peixoto2014hierarchical}. This model is hierarchical, which means that it reveals the community structure of the network at different levels. Level 0 corresponds to the individual nodes, and the higher the level, the lower the number of communities it has. Several communities of a given level can merge into another community in a higher level. The highest level has only one community that corresponds to the whole network.
	
\begin{figure*}
	\centering
	\includegraphics[width=0.7\textwidth]{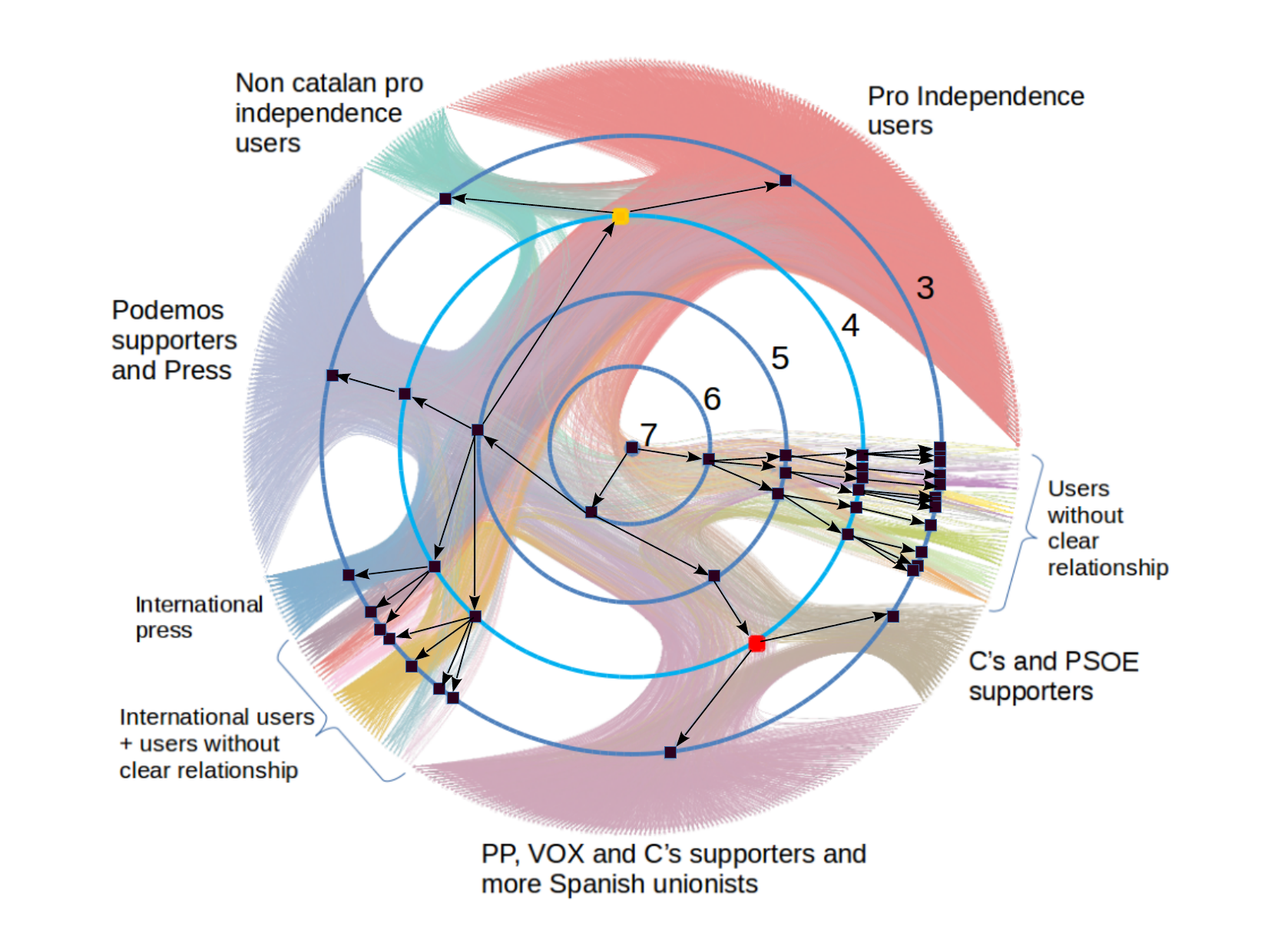}
	\caption{\small Community structure of the retweet network of the Catalan independence Twitter conversation to explain the elite selection. Only the candidates to conform the elite users (pink set of Figure \ref{fig:rectangles}) are shown. Each level is a concentric blue circumference in which each black square represents a community.  The lowest level represented, level 3, is the outermost circumference, and the highest level, level 7, in which there is only one community, it is the center of the circle. The edge of the figure correspond to the single users of the pink rectangle, and the lines between them are the edges (retweets). The two communities of the final elite users are marked with a red and yellow squares.}
	\label{fig:elite}
\end{figure*}

We calculate the community structure with this model for the whole retweet network and analyze the community assignment for the chosen subset of users described above. In Figure \ref{fig:elite} we present the resulting communities for these users, from level 3 to 7. Each level corresponds to a concentric blue circumference in which each black square represents a community.  The lowest level represented, level 3, is the outermost blue circumference, and the highest level, level 7, in which there is only one community, is the center of the circle. The edge of the figure correspond to the chosen subset of users, and the lines between them are the links in the retweet network (retweets). Our goal is to find the highest level where we can find:
	
\begin{itemize}
	\item A community where all the politicians contained belong to political parties that are completely in favor of the Catalan independence: ERC, CUP and PdeCat
	\item A community where all the politicians contained belong to political parties that are completely against of the Catalan independence: Partido Popular (PP), Ciudadanos (C's) and Partido Socialista Obrero Español (PSOE)
\end{itemize}

We have analyzed the users of each community and we have found that the higher level that fulfills these conditions is level 4, where we have marked the two communities with yellow and red squares. The yellow community has 184 users including the politicians of the main political parties in favor of the Catalan independence. We assign them an ideology value of X = +1. The red community has 139 users, including the politicians of the main political parties against independence. We assign them an ideology value of X = -1. These two communities of the 4th level make up the set of elite users. Notice that if we look at the next level (level 5) the against independence community stays the same, but the pro-independence community merges with a third big community that contains Podemos' politicians and press users. 		
	
\subsubsection*{Listeners selection}
\paragraph{}
As we mentioned before, we will consider that opinions propagate through the retweet network from the elite users to the rest of the users, called listeners. Accordingly, the listeners must be "connected" to the elite. This means that there must be a directed path that starts in each listener and ends in at least one elite user. This condition is explained in Supplementary Fig.S1. Every user that satisfies this requirement, along with the set of elite users, will make up the retweet network considered in this study. This final retweet network, shown in Supplementary Fig.S2, is made up of 1639337 users.

\subsection*{Parameters to characterize the opinion distribution}
\paragraph{}
Three parameters are used to characterize the opinion distribution \cite{morales2015measuring}, inspired by the electric dipole moment. These parameters are the normalized pole distance (d), the normalized difference in population sized ($\Delta A$) and the polarization index ($\mu$). The normalized pole distance (d) measures the distance between the gravity centers of the two poles, i.e., how differing the opinions of two sides are. The $\Delta A$ is the difference in population of the negative opinions and the positive ones. In third place, following the definition given in  \cite{morales2015measuring}, we say that a population is perfectly polarized when divided in two groups of the same size with opposite opinions. Accordingly, the polarization index is defined as:
\begin{equation}\label{eq:mu}
\mu=(1- \Delta A)d
\end{equation}

\subsection*{Language inference}
\paragraph{}
The language used by the users who participate in the conversation can be inferred. For this purpose, we use the Python implementation of the library called langdetect \cite{nakatani2010langdetect}. We have tested this method in several Spanish short texts (28666 total words) and Catalan short texts (37399 total words). Both languages have a significantly high precision and recall (Table \ref{tab:langdetect}). Therefore, this way we can detect if tweets were written in Catalan, Spanish or any other language. The user's tweets are the original tweets, and those that she has retweeted or quoted. As we want to know the relationship between Catalan and Spanish tweets, tweets in any other language will not be considered. We define the language index as the average among all the tweets in Spanish or Catalan. That is, the language index $L_i$ of each i user is:
$$
L_i = \frac{\sum_{1}^{n_{i}}l_r}{n_{i}}
$$
where each Catalan tweet have a value of $ l_r = 1 $ and each Spanish tweet a value of $ l_r = -1 $.  $n_{i}$ is the total number of Catalan or Spanish tweets of the user i. Therefore, the language index will have values from -1 to 1. 
\begin{table}
	\centering
	\scalebox{0.9}{ \begin{tabular}{ | c | c | c |}
		\hline
		Language & Precision & Recall \\ \hline 
		Catalan&0.9783 &0.9783\\ \hline 
		Spanish&0.9766 &0.9483\\ \hline 
	\end{tabular}}
	\caption{\label{tab:langdetect} \small Precision and recall of testing the method  \cite{nakatani2010langdetect} to detect Catalan and Spanish languages.}
\end{table}

\section*{Data availability}
\paragraph{}
The data can be requested from the authors through personal communication.

{\small \bibliography{sample}}

\section{Legends}
\paragraph{}
Figure 1. Temporal evolution of the number of tweets published in a Twitter conversation about the Catalan independence issue from 09/11/2017 to 11/04/2017. The three highest activity peaks match the off-line events of: the referendum day (1-0), the Catalonia independence declaration signed and suspended on the same day, and the Unilateral Declaration of Independence (DUI) and the enforcement of the Article 155 of the Spanish Constitution.
\paragraph{}
Figure 2. Left: opinion distribution of all the users and of the users with an activity corresponding to more than 10 tweets in the Catalan independence conversation.  Right: distribution of user activity. Grey area shows the percentage of users that have posted less than 10 tweets in the whole period.
\paragraph{}
Figure 3. Comparison of the effect of activity, influence and engagement on the users' opinion distribution in a Catalan independences Twitter conversation. Results are filtered by activity (blue), retweets received (purple) and days of participation (pink) for three different threshold values increasing from top to bottom panels.
\paragraph{}
Figure 4. Time evolution of the daily distributions of opinion index ($X_i$) for the Catalan independence twitter conversation. Color indicates the number of users that participate each day. The three days framed in red correspond to the three activity peaks observed in Figure \ref{fig:timeseries}.
\paragraph{}
Figure 5. Temporal evolution of the number of users and the normalized pole distance (d), relative population size ($\Delta A$) and the polarization index $\mu$ for the Catalan independence conversation.
\paragraph{}
Figure 6. Distributions of language index (L) for (a) Elite users, where the pro-independence elite users are in blue and the against-independence in orange, (b) Listeners users and (c) Geo-located in Catalonia users.
\paragraph{}
Figure 7. Interplay between language index and opinion index of the users of the Catalan independence Twitter conversation. Color measures the number of users located in the 2D bin. The red x's correspond to Geo-located in Catalonia users.
\paragraph{}
Figure 8. Effect of the activity in the interplay between language index and opinion index. Three different activity thresholds have been used to filter users' activity: 10 tweets (left), 50 tweets (center) and 210 tweets (right). Color indicates the number of users located in each 2D bin. 
\paragraph{}
Figure 9. Retweets received vs. percentage of participation days of users in the Catalan independence Twitter conversation. Colored rectangles represent three subsets of users according to different criteria thresholds. We will work with the pink set, although the outcome is practically the same with the other two sets.
\paragraph{}
Figure 10. Community structure of the retweet network of the Catalan independence Twitter conversation to explain the elite selection. Only the candidates to conform the elite users (pink set of Figure \ref{fig:rectangles}) are shown. Each level is a concentric blue circumference in which each black square represents a community.  The lowest level represented, level 3, is the outermost circumference, and the highest level, level 7, in which there is only one community, it is the center of the circle. The edge of the figure correspond to the single users of the pink rectangle, and the lines between them are the edges (retweets). The two communities of the final elite users are marked with a red and yellow squares.

\paragraph{}
Table 1. Parameters, defined in Data and Methods, that characterize the opinion distributions of Figure \ref{fig: opinion_distribution}.
\paragraph{}
Table 2. Precision and recall of testing the method  \cite{nakatani2010langdetect} to detect Catalan and Spanish languages.

\section*{Acknowledgements}
\paragraph{}
The authors acknowledge support from Projects No. MTM2015-63914-P and No. PGC2018-093854-B-I00 of the Spanish Ministerio de Economía y Competitividad of Spain and Spanish Ministerio de Ciencia Innovación y Universidades of Spain.

\section*{Author contributions statement}
\paragraph{}
J.A.B., S.M.G, J.C.L. and R.M.B. conceived and designed the research.; J.A.B. performed and conducted the computational results; J.A.B., S.M.G, J.C.L. and R.M.B. analyzed the results and wrote and revised the manuscript.

\section*{Additional information}

\textbf{Competing interests}. 

\noindent The authors declare no competing interests.

\end{document}